\documentclass[10pt,twocolumn,superscriptaddress,showpacs,aps,pre]{revtex4-1}
\usepackage{color}
\usepackage{amsmath}
\usepackage{CJK}
\usepackage{psfrag}
\usepackage{graphicx}

\begin{document}

%\begin{CJK*}{GB} { } % Use default fonts from CJK (see below)
\title{Synchronization of coupled active rotators by common noise}
\author{Anastasiya V.\ Dolmatova}
\affiliation{Institute of Continuous Media Mechanics, UB RAS, Akad.\ Korolev Str.\ 1,
 614013 Perm , Russia}
\author{Denis S.\ Goldobin}
\affiliation{Institute of Continuous Media Mechanics, UB RAS, Akad.\ Korolev Str.\ 1,
 614013 Perm , Russia}
\affiliation{Department of Theoretical Physics, Perm State University, Bukirev Str.\ 15,
 Perm 614990, Russia}
\author{Arkady Pikovsky}
\affiliation{Institute for Physics and Astronomy,
University of Potsdam, Karl-Liebknecht-Str.\ 24/25, 14476 Potsdam-Golm, Germany}
\affiliation{Research Institute for Supercomputing, Nizhny Novgorod State University,
Gagarin Av.\ 23, 606950 Nizhny Novgorod, Russia}
\date{\today}

\begin{abstract}
We study the effect of common noise on coupled active rotators. While such a noise always facilitates
synchrony, coupling may be attractive/synchronizing or repulsive/desynchronizing. We develop an analytical
approach based on a transformation to
approximate angle-action variables and averaging over  fast rotations. For identical rotators,
we describe a transition from full to partial synchrony at a critical value of repulsive coupling. For nonidentical
rotators, the most nontrivial effect occurs at moderate repulsive coupling, where a
juxtaposition of phase locking with frequency repulsion (anti-entrainment) is observed. We show that
the frequency repulsion obeys a nontrivial power law.
\end{abstract}
%05.45.Xt	Synchronization; coupled oscillators
%05.40.Ca	Noise
\maketitle
%\end{CJK*}

\section{Introduction}

Synchronization in populations of oscillators is a spectacular effect, important for many areas
of physics (Josephson junction and laser arrays~\cite{Benz-Burroughs-91,*Nixon_etal-13}, electrochemical and electronic
oscillators~\cite{Kiss-Zhai-Hudson-02a,*Temirbayev_etal-12,*Temirbayev_etal-13}) as
well as for many examples from engineering and life sciences (see reviews~\cite{Acebron-etal-05,Pikovsky-Rosenblum-15}).
There are three generic ways to synchronize a population: (i) by a common periodic force,
(ii) by a mutual attractive coupling, and (iii) by common noise. To characterize synchrony,
one uses the notions of phase locking and frequency entrainment. In the first case, when all oscillators
are synchronized to a periodic force, their phases are locked by this force and the frequencies
are entrained by it. In the case of a mutual coupling, exemplified by the famous Kuramoto model
of mean-field coupled oscillators~\cite{Acebron-etal-05,Kuramoto-75},
one needs an attractive coupling to achieve synchrony, which also manifests itself as mutual phase
locking and mutual frequency entrainment. For a repulsive coupling, the phases of the oscillators
disperse and a state with a vanishing mean field sets on, there the oscillators
are essentially non-interacting, their phases are independent and the frequencies are just
the natural ones.  In the case of synchronization by common noise~\cite{Pikovsky-84,*Pikovsky-84a}, the phases
are most of the time locked to some values randomly varying in time in concordance with the noise waveform,
but the frequencies of the oscillators are not shifted---they are identical to the natural ones.

A nontrivial interrelation between phase locking and frequency entrainment
appears under a common action of coupling and common
noise~\cite{GarciaAlvarez_etal-09,*Nagai-Kori-10,*Braun-etal-12,Pimenova_etal-16,Goldobin_etal-17}. If the coupling
is attractive, both factors lead to phase locking, and the coupling additionally
pulls the frequencies together, so that one observes also frequency entrainment, albeit not perfect.
In the case of repulsive coupling, the two factors act in different directions: noise pulls
the phases together while the coupling pushes them apart. As a result, one still observes
that the phases most of the time are close to each other, but the repulsive interaction
produces \textit{frequency anti-entrainment}: the observed frequencies
are more dispersed than the natural ones~\cite{Pimenova_etal-16,Goldobin_etal-17}.

The goal of this paper is to extend the consideration of the effects due to coupling and common noise
to an important class of systems---coupled active
rotators~\cite{Shinomoto-Kuramoto-86a,*Sakaguchi-Shinomoto-Kuramoto-88a}. Each active rotator is described
by an angle $\varphi$, satisfying, in the autonomous overdamped case, the equation
\begin{equation}
\dot\varphi+B\sin\varphi=\Omega\,.
\label{eq:ar}
\end{equation}
Here parameter $\Omega$ is the torque acting on the rotator.
We will consider below only the case $|\Omega|>|B|$, i.e.\ the free rotators are
rotating with frequency $\sqrt{\Omega^2-B^2}$ and are not static. In the  model of
globally coupled active rotators, first studied
by Shinomoto and Kuramoto~\cite{Shinomoto-Kuramoto-86a,*Sakaguchi-Shinomoto-Kuramoto-88a},
one assumes that the coupling is via the
mean field defined as $R e^{i\Phi}=\langle e^{i\varphi}\rangle$:
\begin{equation}
\dot\varphi_j=\Omega_j-B\sin\varphi_j+\mu R\sin(\Phi-\varphi_j)+\sigma\xi(t)\,,
\label{eq:arc}
\end{equation}
where index $j$ denotes units in the population, and $\xi(t)$ is Gaussian noise with $\langle \xi(t)\xi(t')\rangle=2\delta(t-t')$. The nonidentity of units in this model is due to the different torques $\Omega_j$, determining the individual natural frequencies of elements. The last term on the r.h.s.\ of \eqref{eq:arc} describes common white noise~\cite{Sakaguchi-08}.
In many studies one considered a noisy coupled active rotator model,
with independent noise terms acting on all
elements~\cite{Park-Kim-96,*Tessone_etal-07,*Zaks_etal-03,*Sonnenschein-13,*Ionita-Meyer-Ortmans-14}. Such a
noise contributes
to diversity and destroys synchrony. On the contrary,
common noise
facilitates synchrony~\cite{Goldobin-Pikovsky-04,*Goldobin-Pikovsky-05a,*Teramae-Tanaka-04}. The parameter $\mu$
in model \eqref{eq:arc} is the coupling constant:
positive values of $\mu$ describe  attractive coupling, and negative values of $\mu$
correspond to repulsive coupling.
Equation~\eqref{eq:ar} describes also other systems---Josephson
junctions and theta-neurons. However, in these models the coupling is organized
differently~\cite{Marvel-Strogatz-09,*Laing-14,*Okeefe-Strogatz-16,*Luke-Barreto-So-14,*Montbrio-Pazo-Roxin-15}. Hence, our results are not directly applicable to these systems.

The main tool in studying the models of type~\eqref{eq:arc} is the Ott--Antonsen ansatz~\cite{Ott-Antonsen-08},
which yields a closed system of macroscopic equations for the order parameters
$(R,\Phi)$ (for such an analysis of the Kuramoto--Sakaguchi model see~\cite{Goldobin_etal-17}). We present
these equations in Section~\ref{sec:be}. We need, however, to reformulate these macroscopic
equations in terms of order parameters,
more convenient for the analysis---because angles $\varphi_j$ are
not the true phases. In Section~\ref{sec:op} we focus on the statistical properties of the order
parameters. We use both the original exact equations, and the ones averaged over  fast
rotations. We show, that the order parameter does not vanish (even for a strong
repulsive coupling), which indicates a partial synchrony induced by common
noise. For a weak repulsive coupling, the order parameter is quite large, what means that the
rotators almost always form a cluster, i.e.\ their states nearly coincide. This can be described as
phase locking. Properties of the oscillators' frequencies are studied in Section~\ref{sec:fr}.
We show that in the regime of repulsive coupling, the observed frequencies are pushed apart, their
differences are larger than those of the natural frequencies. Moreover, this effect is singular, as
the frequency differences follow nontrivial power laws in dependence on the mismatch of
the natural frequencies.

\section{Basic equations}
\label{sec:be}

\subsection{Formulation in terms of collective variables}
The active rotator model~\eqref{eq:arc} can be written in the form $\dot\varphi_j=\Omega_j(t)+\text{Im}(H(t)e^{-\varphi_j})$.
Thus, in the thermodynamic limit of an infinitely large population,
it allows for an Ott--Antonsen reduction~\cite{Ott-Antonsen-08} to equations for the coarse-grained complex-valued order
parameters for a subpopulation having the torques
in a small range around $\Omega$ $z(\Omega) =\langle
e^{i\varphi}\rangle|_\Omega$ (for brevity we omit the argument in the equations below):
\begin{equation}
\dot{z}=i(\Omega+\sigma\xi(t))z+\frac{\mu}{2}[Z-Z^* z^2]+\frac{B}{2}(1-z^2)\,.
\label{eq1-01}
\end{equation}
The global mean field $Z=Re^{i\Phi}$ can be represented as the average over the distribution of
the torques $Z=\int g(\Omega)\,z\,\mathrm{d}\Omega$. For a Lorentzian distribution
with mean $\Omega_0$ and half-width $\gamma$,
$g(\Omega)=\gamma/[\pi((\Omega-\Omega_0)^2+\gamma^2)]$, the integration under the assumption
of analiticity in the upper half-plane yields $Z=z(\Omega_0+i\gamma)$. This allows obtaining a closed
equation for the global mean field $Z$:
\begin{equation}
\dot Z=(i\Omega_0-\gamma+i\sigma\xi(t))Z+\frac{\mu}{2}Z(1-|Z|^2)+\frac{B}{2}(1-Z^2)\,.
\label{eq1-02}
\end{equation}
The case of identical rotators just corresponds to $\gamma=0$. In the real variables the equations
read:
\begin{equation}
\begin{aligned}
\dot R&=\frac{\mu}{2}R(1-R^2)-\gamma R +\frac{B}{2}(1-R^2)\cos\Phi\,,\\
\dot \Phi&=\Omega_0+\sigma\xi(t)-\frac{B}{2}\frac{1+R^2}{R}\sin\Phi\,.
\end{aligned}
\label{eq:req}
\end{equation}
For $\mu=\gamma=0$ the system dynamics is conservative: here
$\dot{R}=-R\,(\partial{H}/\partial\Phi)$ and $\dot{\Phi}=R\,(\partial{H}/\partial R)$ with function
$H(R,\Phi)=(\Omega_0+\sigma\xi(t))\ln{R}-(B/2)(R^{-1}-R)\sin\Phi$.
The ``singularity'' at $R=0$ in the equation for the phase in system~\eqref{eq:req} is
due to the uncertainty of the phase at $R=0$, and does not result in any singularities for the
behavior of the order parameter. This follows also from the absence of any singularity
in Eq.~\eqref{eq1-02} for the complex order parameter.

It is convenient to introduce a new order parameter $J=R^2(1-R^2)^{-1}$. In terms of the variables
$(J,\Phi)$ we obtain
\begin{align}
\dot{J}&=\mu J-2\gamma J(1+J)+B\sqrt{J(1+J)}\cos\Phi\,,\label{eq1-04}\\
\dot{\Phi}&=\Omega_0-B\frac{J+1/2}{\sqrt{J(1+J)}}\sin\Phi+\sigma\xi(t)\,.\label{eq1-05}
\end{align}
The new order parameter $J$ varies in the range $0\leq J<\infty$, and the situation of full
synchrony corresponds to $J\to\infty$.
We complement this equation with the dynamics of the angle difference
$\vartheta_\omega=\varphi-\Phi$ between the rotator having the
natural torque $\Omega=\Omega_0+\omega$ and the mean field:
\begin{equation}
\begin{aligned}
\dot{\vartheta}&=\omega-\mu\sqrt{\frac{J}{1+J}}\sin\vartheta\\
&\qquad{}
-B\left[\sin(\Phi+\vartheta)-\frac{J+1/2}{\sqrt{J(1+J)}}\sin\Phi\right].
\label{eq1-06}
\end{aligned}
\end{equation}
Again, for brevity we omit the index at $\vartheta$.
Equations~\eqref{eq1-04}--\eqref{eq1-06} is the basic system to be analyzed below.

\subsection{Natural variables for quantifying regimes close to synchrony}
While the order parameters $R=|Z|$ and $J$  are natural quantifiers for characterization
of the order in the angles of rotators, the argument of the complex mean field $\Phi$
is not the proper oscillation phase, as it rotates non-uniformly. This inhomogeneity
is ``inherited'' from the angle $\varphi$: this variable is not the true ``phase''
which should rotate
uniformly for a single rotator. This non-uniformity of the rotations results in a non-zero value for
the order parameters $R,J$ even in an uncoupled, not forced population.
Thus, it is convenient to introduce new phase variables and to ``correct'' correspondingly
the order parameter $J$.

It is instructive to mention that in the case $\mu=\gamma=0$, Eqs.~\eqref{eq1-04}--\eqref{eq1-05}
can be written as Hamilton equations with the Hamilton function $H(J,\Phi)=(\Omega_0+\sigma\xi(t)) J-
B\sqrt{J(1+J)}\sin\Phi$.
The proper transformation would be a transformation to the action--angle variables for this Hamiltonian.
This, however, results in cumbersome, not tractable expressions. Therefore we perform a transformation
to  the action--angle variables of the Hamiltonian $\widetilde{H}=\Omega_0 J-B J\sin\Phi$,
which is a good approximation
to $H$ for large $J$ (i.e. for regimes close to synchrony). The new variables $(I,\Psi)$ are expressed as
\begin{equation}
\sin\Phi=\frac{a-\cos\Psi}{1-a\cos\Psi},\quad J= I(1-a\cos\Psi),\quad a=\frac{B}{\Omega_0}\;.
\label{eq:ct}
\end{equation}
The exact equation for the new order parameter $I$ reads
\begin{align}
\dot{I}=&(\mu-2\gamma)I-2\gamma(1-a\cos\Psi)I^2
\nonumber\\%[5pt]
&+J_0\frac{\sqrt{1-a^2}\sin\Psi}{1-a\cos\Psi}I\Bigg[
 {\textstyle \sqrt{1+\frac{1}{J}}-1}
\nonumber\\%[5pt]
&+\frac{a}{1-a^2}(a-\cos\Psi)\bigg({
 \textstyle\sqrt{1+\frac{1}{4J(J+1)}}-1}
 \bigg)\Bigg]
\nonumber\\%[5pt]
&
 -\frac{a}{\sqrt{1-a^2}}I\sin\Psi\sigma\xi(t)\,.
\label{eq1-15}
\end{align}
The equation for $\Psi$ takes the form
\begin{align}
\dot{\Psi}=&\nu_0-\frac{\Omega_0a}{\sqrt{1-a^2}}(a-\cos\Psi)
 \bigg({\textstyle\sqrt{1+\frac{1}{4J(J+1)}}-1}\bigg)
\nonumber\\%[5pt]
&
+\frac{\sigma}{\sqrt{1-a^2}}(1-a\cos\Psi)\,\xi(t)\,,
\label{eq1-16}
\end{align}
where  $\nu_0=\sqrt{\Omega_0^2-B^2}$ is the ``true frequency'' of an active rotator with torque $\Omega_0$.
For the transformation of the angles of the rotators to the phases, $\varphi\to\psi$, we use not their natural torques $\Omega$, but
the mean one $\Omega_0$. Thus the transformation looks exactly like \eqref{eq:ct}:
\[
\sin\varphi=\frac{a-\cos\psi}{1-a\cos\psi}\;,
\]
and the equation for the oscillator having the torque  $\Omega=\Omega_0+\omega$ reads
\[
\begin{aligned}
\dot{\psi} =& \nu_0 +\omega\frac{(1-a\cos\psi)}{\sqrt{1-a^2}}\\
& + \frac{\mu}{\sqrt{1-a^2}}(1-a\cos\psi)\sqrt{\frac{J}{1+J}}\sin(\Phi-\varphi)\\
& + \frac{\sigma}{\sqrt{1-a^2}}(1-a\cos\psi)\,\xi(t)\, .
\end{aligned}
\]
Denoting the normalized deviation to the mean torque as $\nu=\omega/\sqrt{1-a^2}$, we
obtain the equation for the phase difference $\theta=\psi-\Psi$ in the form
\begin{equation}
\begin{aligned}
\dot\theta=&\nu\big(1-a\cos(\Psi+\theta)\big)
\\
&
 +\frac{\Omega_0a\bigg({\textstyle\sqrt{1+\frac{1}{4J(J+1)}}-1}\bigg)}{\sqrt{1-a^2}}(a-\cos\Psi)
 \\
&
 +\frac{\mu\Big(a\big(\sin(\Psi+\theta)-\sin\Psi\big)-\sin\theta\Big)}{1-a\cos\Psi}\left[1+\frac{1}{J}\right]^{-1/2}
\\
&
 + \frac{a\sigma}{\sqrt{1-a^2}}\Big(\cos\Psi-\cos(\Psi+\theta)\Big)\xi(t)\,.
\end{aligned}
 \label{eq1-17}
\end{equation}
Eqs.~(\ref{eq1-15})--(\ref{eq1-17}) for $\dot{I}$, $\dot{\Psi}$ and $\dot{\theta}$ are exact. However,
their essential advantage to the original equations
\eqref{eq1-04}--\eqref{eq1-06} is for regimes close to synchrony, where $I,J\gg 1$.
In this limit, many terms in Eqs.~(\ref{eq1-15})--(\ref{eq1-17}) vanish
and we obtain the following tractable system:
\begin{align}
\dot{I}=&(\mu-2\gamma)I-2\gamma(1-a\cos\Psi)I^2
 -\frac{aI\sin\Psi}{\sqrt{1-a^2}}\sigma\xi(t)\,,
\label{eq:I}\\
\dot{\Psi}=&\nu_0
+\frac{\sigma}{\sqrt{1-a^2}}(1-a\cos\Psi)\,\xi(t)\,,
\label{eq:P}\\
\dot\theta=&\nu\big(1-a\cos(\Psi+\theta)\big)
\nonumber \\
&
 +\frac{\mu\Big(a\big(\sin(\Psi+\theta)-\sin\Psi\big)-\sin\theta\Big)}{1-a\cos\Psi}
\nonumber \\
&
 + \frac{a\sigma\Big(\cos\Psi-\cos(\Psi+\theta)\Big)}{\sqrt{1-a^2}}\xi(t)\,.
 \label{eq:th}
\end{align}
The system of equations \eqref{eq:I}--\eqref{eq:th} is a skew system, where the
variable $\theta$ depends on the dynamics of $\Psi$, but not vice versa.
To determine the statistical properties of the order parameter, it is sufficient
to study first two equations \eqref{eq:I} and \eqref{eq:P}; to find the statistics of the units in
the population, one has to add Eq.~\eqref{eq:th}.

The system of stochastic differential equations \eqref{eq:I}--\eqref{eq:th} yields a Fokker--Planck
equation for the probability density $W(I,\Psi,\theta,t)$. Even if one confines to the properties of the order parameter, one has to analyze the density depending on two variables $(I,\Psi)$, which is
hardly possible. However, in the case of fast oscillations, the phase $\Psi$ is a fast variable, and
one can average the Fokker--Planck equation over these fast oscillations. As a result, only
the variables $(I,\theta)$ remain; moreover, the equations for these variables decouple. We present
the details of the derivation in Appendix~\ref{sec:apa}. The resulting averaged stochastic differential equations read:
\begin{align}
\dot{I}&=(\mu-2\gamma)I -2\gamma I^2 +\widetilde{\sigma}^2I
 -\widetilde{\sigma}\,I\zeta_1(t)\,,
\label{eq:Iav}
\\[10pt]
\dot{\theta}&=\nu -
\big(\mu +\widetilde{\sigma}^2\big)\sin\theta
\nonumber\\
&\qquad
+\widetilde{\sigma}\sin\theta\zeta_1(t)
 -\widetilde{\sigma}(1-\cos\theta)\zeta_2(t)\,,
\label{eq:thav}
\end{align}
where we introduce the normalized noise amplitude
\[
\widetilde{\sigma}\equiv\frac{a\,\sigma}{\sqrt{2(1-a^2)}}\,.
\]
The averaged equations contain two effective independent white noise terms $\zeta_{1,2}(t)$.

\section{Statistical properties of the order parameter}
\label{sec:op}
\subsection{Identical oscillators --- synchronous state stability}
We start the analysis of the dynamics of the population of coupled active rotators
under common noise with the case of identical oscillators, $\gamma=0$.
Here the equation for the order parameter $I$~\eqref{eq:Iav} can be recast as
\begin{equation}
\frac{\mathrm{d}}{\mathrm{d}t}\ln I=\mu+\widetilde{\sigma}^2-\widetilde{\sigma}\zeta_1(t)\,.
\label{eq:strI}
\end{equation}
Averaging this equation, one can find the Lyapunov exponent $\lambda\equiv\langle(\mathrm{d}/\mathrm{d}t)\ln{I}\rangle$ determining the exponential growth/decay of $I$;
\begin{equation}
\lambda=\mu+\widetilde{\sigma}^2\;.
\label{eq:aprl}
\end{equation}
Thus, the population synchronizes, $I\to\infty$, for $\lambda>0$, i.e.\ if the coupling is attractive or not too large repulsive
\begin{equation}
\mu>\mu_c=-\widetilde{\sigma}^2\;.
\label{eq:thr}
\end{equation}

%%%%%%%%%%%%%%%%%%%%%%%%%%%%%%%%%%%%%%%%%%%%%%%%%%%%%%%%%
\begin{figure}[!thb]
\centerline{
\includegraphics[width=0.9\columnwidth]%
 {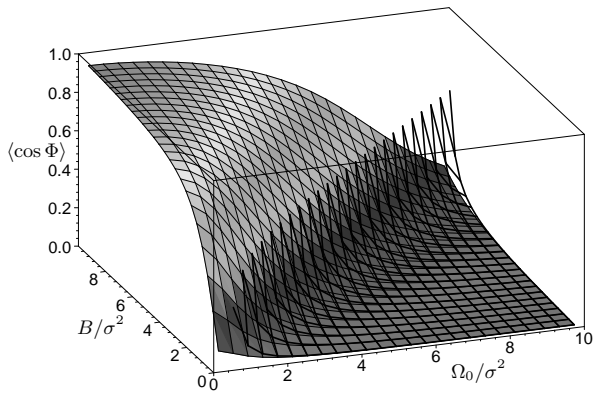}}
\caption{ The average value $\langle\cos\Phi\rangle$ determining the Lyapunov
exponent for the order parameter $I$ (or $J$) [see Eqs.~(\ref{eq3-08}) and (\ref{eq3-07})] is plotted
as a function of the system parameters (filled surface). The asymptotic
behavior  $\langle\cos\Phi\rangle\approx0.5 B\sigma^2/(\Omega_0^2-B^2)$ corresponding
to Eq.~(\ref{eq:aprl}) (wireframe) is compared against the exact formula~(\ref{eq3-08}).}
  \label{fig1}
\end{figure}
%%%%%%%%%%%%%%%%%%%%%%%%%%%%%%%%%%%%%%%%%%%%%%%%%%%%%%%%%

The synchronization threshold can be also
determined without the averaging, for general parameters of the system. Indeed, taking the limit $J\gg 1$ for $\gamma=0$ in
Eqs.~\eqref{eq1-04}--\eqref{eq1-05}, we obtain:
\begin{align}
\frac{\mathrm{d}}{\mathrm{d}t}\ln J &=\mu + B \cos\Phi\,,
\label{eq3-05}
\\[5pt]
\dot{\Phi}&=\Omega_0- B \sin\Phi+\sigma\xi(t)\,.
\label{eq3-06}
\end{align}
From Eq.~(\ref{eq3-05}), the Lyapunov exponent governing the growth of $J$ can be expressed as
\begin{equation}
\lambda=\mu+B \langle\cos\Phi\rangle\,.
\label{eq3-07}
\end{equation}
On the other hand, because Eq.~(\ref{eq3-06}) is independent of $J$, the statistics of $\Phi$ follows from the corresponding Fokker--Planck equation, written for the probability density of $\Phi$.
The stationary solution of this equation reads
\[
\rho(\Phi)=\frac{\nu_\sigma}{2\pi(1-\mathrm{e}^{-2\pi\Omega_0})\sigma^2}
\int_\Phi^{\Phi+2\pi}\mathrm{d}\Phi_1\mathrm{e}^{U(\Phi)-U(\Phi_1)}\,,
\]
where $U(\Phi)\equiv(\Omega_0/\sigma^2)\Phi+(B/\sigma^2)\cos\Phi$, and the average frequency $\nu_\sigma$ is to be determined from the normalization condition $\int_0^{2\pi}\rho(\Phi)\,\mathrm{d}\Phi=1$. Thus,
\begin{equation}
\langle\cos\Phi\rangle=\frac{
\displaystyle\int_0^{2\pi}\mathrm{d}\Phi\,\cos\Phi
 \int_\Phi^{\Phi+2\pi}\mathrm{d}\Phi_1\mathrm{e}^{U(\Phi)-U(\Phi_1)}}
{\displaystyle\int_0^{2\pi}\mathrm{d}\Phi
 \int_\Phi^{\Phi+2\pi}\mathrm{d}\Phi_1\mathrm{e}^{U(\Phi)-U(\Phi_1)}}\,.
\label{eq3-08}
\end{equation}
The results of calculations with this expression are compared in Fig.~\ref{fig1} with the approximate
formula for large oscillation frequencies~\eqref{eq:strI}, which
corresponds to $\langle\cos\Phi\rangle=\widetilde{\sigma}^2/B=0.5B\sigma^2/(\Omega_0^2-B^2)$.

In the domain where  $\lambda>0$, the synchronous state $R=1$, $I=J=\infty$ is an adsorbing one:
starting from any initial conditions, the synchronous state sets on.
While from Eq.~\eqref{eq:strI} one could conclude that for a strong repulsive coupling,
where $\mu<-\widetilde{\sigma}^2$ and $\lambda<0$, the order parameter $I$ tends to zero;
we have to remind that this equation is valid for large values of $I$ only. For asynchronous states with small $R$, one has to study full equations \eqref{eq:req}, which show that the order parameter $R$
never vanishes exactly. Unfortunately, an analytic exploration of the two-dimensional stochastic system
\eqref{eq:req} [or, equivalently, of \eqref{eq1-04}--\eqref{eq1-05}] is hardly possible, thus we studied it numerically
and present the results together with those for nonidentical oscillators in the next section.

%%%%%%%%%%%%%%%%%%%%%%%%%%%%%%%%%%%%%%%%%%%%%%%%%%%%%%%%%
\begin{figure}[!thb]
\centerline{
\includegraphics[width=0.44\textwidth]%
 {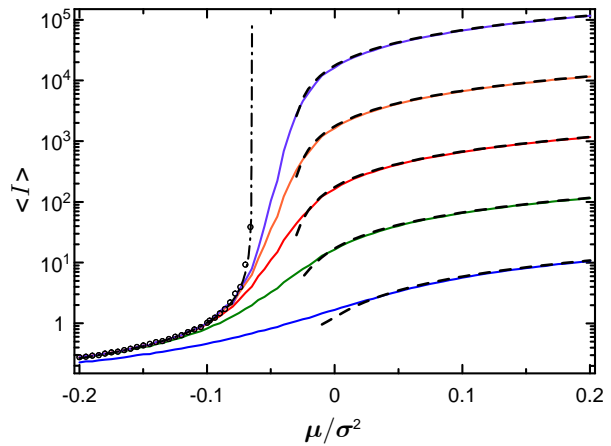}}
\caption{(Color online) The time-average value of the order parameter $\langle I\rangle$ for
an ensemble of nonidentical oscillators {\it vs} coupling strength is plotted
for $\Omega_0=10$, $J_0=2.5$, and values of the frequency band
half-width $\gamma/\sigma^2=0$ (circles),
$10^{-6}$, $10^{-5}$, $10^{-4}$, $10^{-3}$, $10^{-2}$ (solid lines, from top to bottom). Solid lines
represents the results of numerical simulation for Eqs.~(\ref{eq1-04})--(\ref{eq1-05}) with $I$
determined by Eq.~(\ref{eq:ct}). The analytical estimates (\ref{eq3-10}) plotted with dashed lines
appear to be in a good agreement with the results of numerical simulations.}
  \label{fig2}
\end{figure}
%%%%%%%%%%%%%%%%%%%%%%%%%%%%%%%%%%%%%%%%%%%%%%%%%%%%%%%%%

\subsection{Nonidentical oscillators}
The perfect synchrony becomes impossible for an ensemble of nonidentical
oscillators ($\gamma >0$), and
the order parameters fluctuate in a finite range for  all values of the parameters of the
system. An analytical description is possible for the averaged stochastic equation~\eqref{eq:Iav},
valid close to synchrony. We can rewrite it as
\begin{align}
\frac{\mathrm{d}}{\mathrm{d}t}\ln{I}&=\lambda-2\gamma -2\gamma I
-\widetilde{\sigma}\,\zeta_1(t)\,.
\label{eq3-09}
\end{align}
For a stochastically stationary regime, the average of the time derivative should vanish, thus we immediately
find the average value of the order parameter $I$:
\begin{align}
\langle{I}\rangle&=\frac{\lambda}{2\gamma}-1\,.
\label{eq3-10}
\end{align}
In Fig.~\ref{fig2}, one can see a good agreement between this formula and the results of numerical
simulation of full equations \eqref{eq:req}. Here we also present the values of $\langle I\rangle$ for
the case of identical oscillators $\gamma=0$ (obtained via direct numerical simulations of the original
stochastic equations).

Furthermore, stochastic equation (\ref{eq:Iav}) yields the following
Fokker--Planck equation for the probability density $\rho(I)$;
\begin{align}
&\frac{\partial\rho}{\partial t}
+\frac{\partial}{\partial I}\Big[\big(\lambda I-2\gamma I(I\!+\!1)\big)\rho\Big]
 -\widetilde{\sigma}^2
\frac{\partial}{\partial I}\!\left(\!I\frac{\partial}{\partial I}\left(I\rho\right)\!\right)=0\,.
\nonumber
\end{align}
The stationary solution to this equation reads
\[
\rho(I)=\frac{\displaystyle
 I^\frac{\mu-2\gamma}{\widetilde{\sigma}^2}\mathrm{e}^{-\frac{2\gamma}{\widetilde{\sigma}^2}I}}
 {\displaystyle
 \left(\frac{\widetilde{\sigma}^2}{2\gamma}\right)^{\frac{\mu-2\gamma}{\widetilde{\sigma}^2}+1}
 \Gamma\left(\frac{\mu-2\gamma}{\widetilde{\sigma}^2}+1\right)}\,,
\]
where $\Gamma(\cdot)$ is the Gamma function. This expression
allows finding also the higher moments $\langle I^m\rangle$. As mentioned above, this formula is inaccurate for small $I$;
therefore, it cannot be used for calculation of
those statistical characteristics of $I$ for which the contribution of small values is significant, even if
the average $\langle I\rangle$ is large.

\section{Phase dynamics for nonidentical oscillators}
\label{sec:fr}

\subsection{Frequency entrainment and anti-entrainment}

\begin{figure}[!thb]
\centering
\includegraphics[width=0.99\columnwidth]%
 {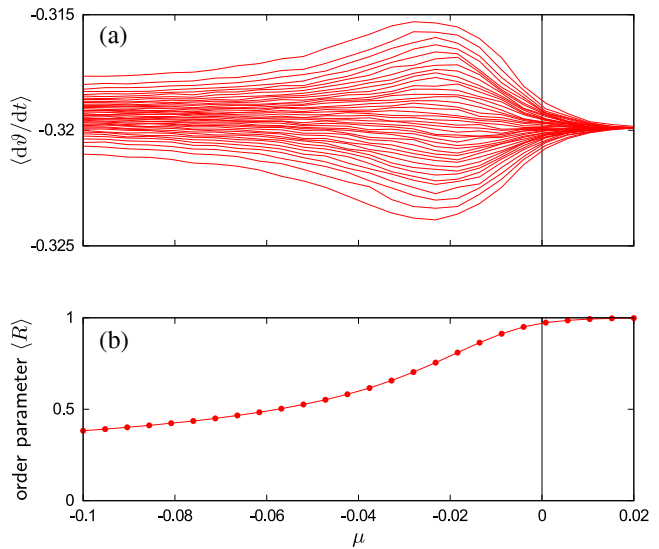}
\caption{(Color online) (a):~Frequencies in an ensemble of 41 active rotators with a Gaussian distribution of torque
parameters $\Omega$, in dependence on the coupling strength $\mu$. (b):~The averaged order parameter $\langle R\rangle$.
Parameters of the model: $\Omega_0=10$, $B=2.5$, $\sigma^2=0.5$, the standard deviation of the distribution of torques
$5\cdot 10^{-4}$. One can see the repulsion of the frequencies in the whole range of
negative values of $\mu$, with a maximal effect around $\mu=-0.022$.}
\label{fig3}
\end{figure}

Let us consider the effect of the interplay of common noise and coupling on the individual mean
frequencies of rotators $\langle \dot\varphi\rangle$.
These frequencies in the absence of coupling ($\mu=0$) are just the natural frequencies
$\sqrt{\Omega^2-B^2}$. In the presence of a coupling, the phases of the rotators are either attracted
to each other (for $\mu>0$) or are repelled (for $\mu<0$). On the contradistinction,
common noise always brings the phases together, resulting in a non-zero (or even large) value of the order
parameter. For an attractive coupling, the latter pulls the frequencies together and the frequency
differences are smaller than in the uncoupled case---this is the usual
situation of frequency entrainment. For a repulsive coupling, the repelling of the phases due to the coupling
leads to the repulsion of the frequencies, and their differences become larger than for the uncoupled
case---this can be called frequency anti-entrainment, see Fig.~\ref{fig3}. This effect is not
present for the case
of a repulsive coupling without noise, because then the phases are just distributed uniformly
so that the mean field vanishes and no effect of repulsion is observed.

%%%%%%%%%%%%%%%%%%%%%%%%%%%%%%%%%%%%%%%%%%%%%%%%%%%%%%%%%

\begin{figure}[!thb]
\centerline{
\includegraphics[width=0.77\columnwidth]%
 {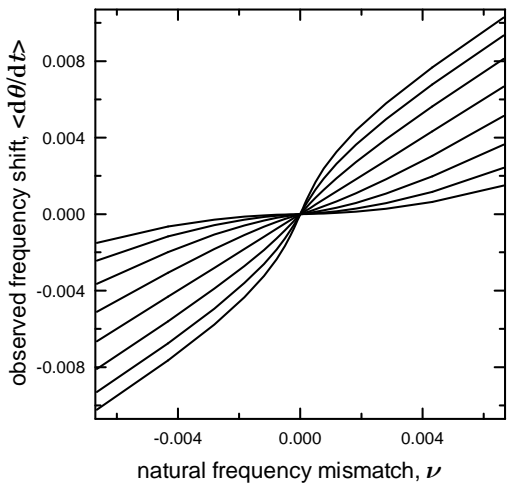}\quad}
\caption{The average frequency shift is plotted {\it vs} the natural frequency
mismatch for $B=2.5$, $\Omega_0=10$, $\sigma^2=0.5$, $\gamma=5\cdot10^{-4}$,
$\mu=0.02,\,0.015,\,0.01,\,0.005,\,0,\,-0.005,\,-0.01,\,-0.015$ (from bottom
to top on the r.h.s.). The frequencies are obtained by virtue of direct numerical simulation
of Eqs.\,(\ref{eq1-04})--(\ref{eq1-06}).
}
  \label{fig4}
\end{figure}

In Fig.~\ref{fig3} we presented the simulations demonstrating entrainment and anti-entrainment
for a finite, in fact relatively small population
of the rotators. The theory above is valid, however, in the thermodynamic limit. To illustrate the effect
of the frequencies anti-entrainment in this limit, and to compare with the theory below, we simulated a set
of one equation for the mean field  \eqref{eq1-02} and of several equations \eqref{eq1-06},
with different values of the natural torque parameter $\omega$. The observed frequencies for
the elements of population \eqref{eq1-06} are plotted in Fig.~\ref{fig4} {\it vs} the natural frequencies
$\sqrt{(\Omega_0+\omega)^2-B^2}$. One can see that in the absence of the coupling ($\mu=0$), the observed
frequencies are just the natural ones, while the effects of entrainment and of anti-entrainment
are evident for the attractive $\mu>0$ and the repulsive $\mu<0$ couplings, respectively.

%%%%%%%%%%%%%%%%%%%%%%%%%%%%%%%%%%%%%%%%%%%%%%%%%%%%%%%%%
%%%%%%%%%%%%%%%%%%%%%%%%%%%%%%%%%%%%%%%%%%%%%%%%%%%%%%%%%

A quantitative description of the discussed effect requires a statistical evaluation of the
dynamics of the phase differences $\theta$. This is possible close to synchrony $I\to\infty$, where
the dynamics of $\theta$ obeys Eq.~(\ref{eq:thav}).
For the probability density $\overline{w}(\theta,t)$ we can write the Fokker--Planck equation
\begin{equation}
\begin{gathered}
\frac{\partial}{\partial t}\overline{w}(\theta,t)+\frac{\partial}{\partial \theta}
\left[\big(\nu-(\mu+\widetilde{\sigma}^2)\big)\overline{w}(\theta,t)\right]=\\
\widetilde{\sigma}^2\Big[\frac{\partial}{\partial \theta}\sin\theta\frac{\partial}{\partial \theta}\sin\theta \overline{w}(\theta,t)\\
+\frac{\partial}{\partial \theta}(1-\cos\theta)\frac{\partial}{\partial \theta}(1-\cos\theta) \overline{w}(\theta,t)\Big]\;.
\end{gathered}
\label{eq:thfpns}
\end{equation}
We look for a stationary solution of this equation with some flux $q$. This flux is related to the
mean frequency as $\langle\dot \theta\rangle=2\pi q$, so the stationary solution of Eq.~\eqref{eq:thfpns}
fulfills the following ODE
\begin{equation}
\frac{\langle\dot{\theta}\rangle}{2\pi}=q=(\nu -\mu\sin\theta)\,\overline{w}(\theta)
-2\widetilde{\sigma}^2\frac{\mathrm{d}}{\mathrm{d}\theta}\Big((1-\cos\theta)\,\overline{w}(\theta)\Big)\,.
\label{eq4-02}
\end{equation}
 From Eq.~(\ref{eq4-02}), one can express
 $\overline{w}(\theta)$ and employ the normalization condition $\int_0^{2\pi}\overline{w}(\theta)\,\mathrm{d}\theta=1$ to obtain
\begin{align}
\langle\dot{\theta}\rangle=4\pi\widetilde{\sigma}^2\Bigg\{
\int\limits_0^{2\pi}\mathrm{d}\theta
\int\limits_\theta^{2\pi}\mathrm{d}\psi
 \frac{(1-\cos\psi)^{\mu/(2\widetilde{\sigma}^2)}}
  {(1-\cos\theta)^{1+\mu/(2\widetilde{\sigma}^2)}}
\qquad\nonumber\\%[5pt]
\times\exp\left[-\frac{\nu}{2\widetilde{\sigma}^2}\left(\cot\frac{\theta}{2}-\cot\frac{\psi}{2}\right)\right]\Bigg\}^{-1}.
\label{eq4-03}
\end{align}

\begin{figure}[!thb]
\centering
\includegraphics[width=0.77\columnwidth]%
 {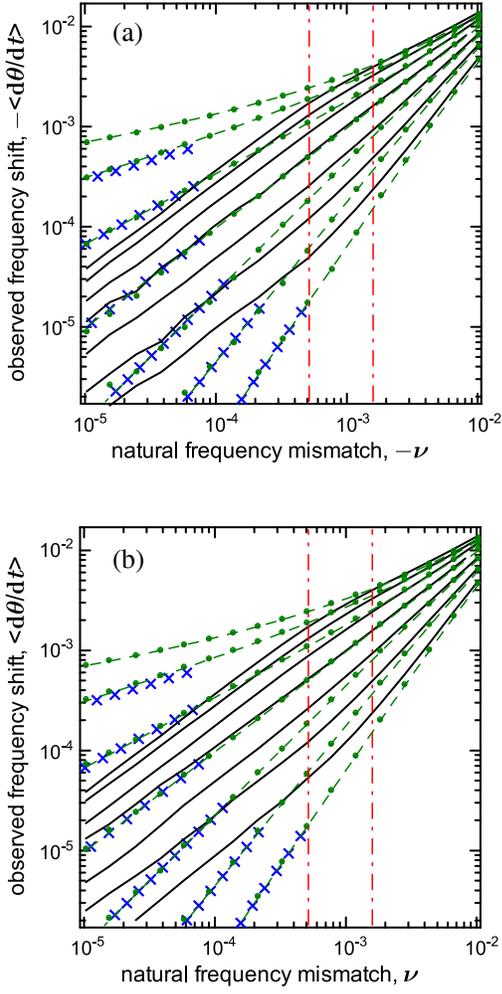}%\\[10pt]
\caption{(Color online) The average frequency shift is plotted on a logarithmic scale
{\it vs} the natural frequency
mismatch for $B=2.5$, $\Omega_0=10$, $\sigma^2=0.5$,
$\mu=0.015$, $0.01$, $0.005$, $0$, $-0.005$, $-0.01$, $-0.015$ (from bottom to top). Black solid
lines: the results of direct numerical simulation for stochastic
system~(\ref{eq1-04})--(\ref{eq1-06}) with $\gamma=5\cdot10^{-4}$; green circles:
Eqs.~(\ref{eq1-04})--(\ref{eq1-06}) with $\gamma=0$; green dashed lines:
analytical theory~(\ref{eq4-03}); blue crests: asymptotic law~(\ref{eq:afr}). The results are
provided for both negative (a) and positive (b) values of $\nu$ as these cases are not identical.
Deviations of the simulated frequencies from the asymptotic law are due to finiteness of the
order parameter $I$ in simulations, while at the derivation of (\ref{eq4-03}), (\ref{eq:afr})
we assumed $I\to\infty$.
% [even
% though at small values of $\gamma/\Omega_0$, for the results of the direct numerical simulation,
% the discrepancy is small against the background of the numeric inaccuracy, and the accuracy of
% the analytical theory does not allow to distinguish these cases].
Red vertical dash-dotted lines: the
natural torque shifts $\omega=\gamma$ and $\omega=3.08\gamma$, half of the rotators in the
ensemble with the Lorentzian distribution of torques
have $|\omega|>\gamma$, and 20\% --- $|\omega|>3.08\gamma$. With these lines one can
see that the approximation $I\to\infty$, adopted at the derivation of
the analytical expressions (\ref{eq4-03}), (\ref{eq:afr}), which neglects the finiteness of $I$ and misses the transition to a linear
law for $\omega\to 0$ (can be seen for the black solid lines), is relevant for a considerable
fraction of the rotators in the ensemble.
}
  \label{fig5}
\end{figure}

A remarkable feature of the exact expression for the frequency of oscillators is
its singular behavior for small natural frequency differences $\nu$. Referring  to Appendix
\ref{sec:apb} for detailed calculations, we present here the resulting power-law dependence:
\begin{align}
\langle\dot{\theta}\rangle&\approx
\frac{2\sqrt{\pi}\,\Gamma\big(|m+\frac12|+\frac12\big)\,\widetilde{\sigma}^2}
 {\Gamma\big(|m+\frac12|\big)\,\Gamma\big(|2m+1|\big)}
 \left(\frac{\nu}{2\widetilde{\sigma}^2}\right)^{|2m+1|}\,.
\label{eq:afr}
\end{align}
Eq.~(\ref{eq:afr}) describes the asymptotic law for $\nu/\widetilde{\sigma}^2\ll1$, which is
valid if the synchrony is high $I\to\infty$. We compare these analytic results with numerical
simulations in Fig.~\ref{fig5}.

%%%%%%%%%%%%%%%%%%%%%%%%%%%%%%%%%%%%%%%%%%%%%%%%%%%%%%%%%

\begin{figure}[!t]
\centering
\includegraphics[width=0.97\columnwidth]%
 {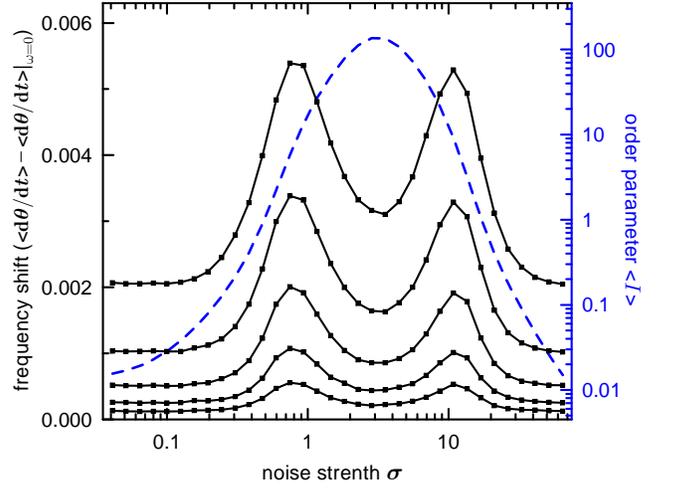}
\caption{(Color online)
The average frequency shift (black solid lines) and order parameter $\langle{I}\rangle$ (blue dashed line) are plotted {\it vs} common noise strength $\sigma$ for an ensemble with fixed inherent parameters $B=2.5$, $\Omega_0=10$, $\mu=-0.02$, and $\gamma=5\cdot10^{-4}$: the results of numerical simulation of stochastic system~(\ref{eq1-04})--(\ref{eq1-06}) for $\omega/\gamma=0.25$, $0.5$, $1$, $2$, $4$.
}
  \label{fig6}
\end{figure}

Above we employed the
derived analytical formulae and the results of numerical simulation in Figs.~\ref{fig3}--\ref{fig5}, to present
a comprehensive picture of the phenomenon of frequency repulsion for repulsive couplings: we fixed the strength
of noise and considered the coupling constant as the major parameter.
It is, however, instructive to show the dependence of the order parameter and of the average frequencies on the
intensity of the external common noise, for fixed parameters of the ensemble and of the coupling.
In Fig.~\ref{fig6}, one can see that the effect of frequency repulsion is maximal for a moderate noise and
becomes again small for a strong noise. The reason is, that for a strong common noise, the phase rotation [see
Eq.~\eqref{eq1-05}] becomes more homogeneous and the contribution of the common noise to the Lyapunov
exponent~\eqref{eq3-07}, which is $\propto\langle\cos\Phi\rangle$, diminishes. This leads
to a decrease of the mean field $\langle{I}\rangle$, and for small mean fields the
dispersing action of repulsive coupling is weaker. Indeed, one can consider the
system~\eqref{eq1-04}--\eqref{eq1-05} for the case of large $\Omega_0$ and $\sigma^2$, and for
moderate $B$ (see Appendix~\ref{sec:apc}) and, similarly to Eq.~\eqref{eq3-10}, derive an equation for the
order parameter
\begin{equation}
\langle{I}\rangle=\frac{1}{2\gamma}\left(\mu +\frac{B^2\sigma^2}{2(\Omega_0^2+\sigma^4)}\left(1+\frac{1}{2}\langle{I^{-1}}\rangle\right)\right)-1\,.
\label{eq:lnoise}
\end{equation}
In this equation, the term $\propto B^2$ is small for both small and large values of $\sigma$. Thus, the synchronization and the frequency repulsion effects in the system under consideration are most pronounced for a moderate common noise strength.
It appears that the extrema of the dependencies in Fig.~\ref{fig6} should not be interpreted as a sort of resonant behavior, since they are not related to any specific matching of several time scales in the system (as, e.g., for
stochastic or coherence resonances~\cite{Benzi-Sutera-Vulpiani-81,*Gang-etal-93,*Gammaitoni-etal-98,Pikovsky-Kurths-97}). These maxima are rather 
related to a non-monotonous dependence of the order parameter
on the noise intensity.
%; however, apart from these dependencies, one cannot exclude some other resonant phenomena in the system dynamics.

%%%%%%%%%%%%%%%%%%%%%%%%%%%%%%%%%%%%%%%%%%%%%%%%%%%%%%%%%
%%%%%%%%%%%%%%%%%%%%%%%%%%%%%%%%%%%%%%%%%%%%%%%%%%%%%%%%%

\subsection{Phase difference slips}

\begin{figure}[!t]
\centering
\includegraphics[width=0.98\columnwidth]%
 {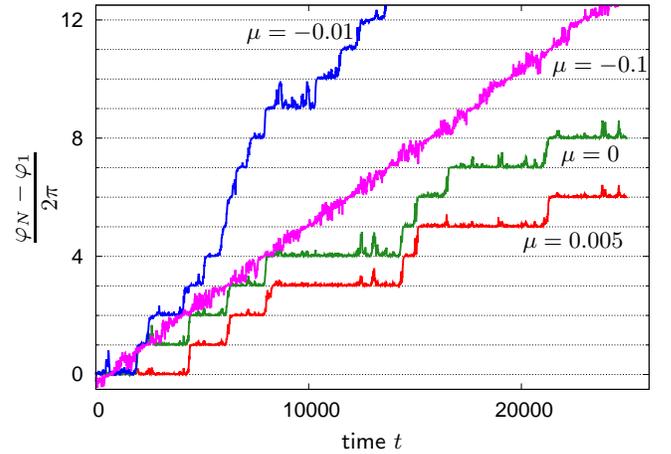}
\caption{(Color online)
The dynamics of the angle differences $\varphi_N(t)-\varphi_1(t)$ between the most fast and most slow rotators
in a population illustrated in Fig.~\ref{fig3}. Green: uncoupled rotators $\mu=0$; red: attractive
coupling $\mu=0.005$; blue: weak repulsive coupling $\mu=-0.01$; magenta: strong repulsive coupling $\mu=-0.1$.
Only in the latter case there are no phase slips, in all other cases one can clearly see long epochs where
$\varphi_N(t)-\varphi_1(t)\approx 2\pi m$.
}
  \label{fig7}
\end{figure}

The effect of frequency anti-entrainment, demonstrated above numerically and described analytically,
appears at first glance counter-intuitive. Indeed, it is observed in regimes with strong phase locking,
where the order parameter is large. This means that most of the time the rotators stay together.
For a usual synchronization by an attractive coupling,  the phase locking and
the frequency entrainment come together. An independence of the phase locking from the frequency entrainment is, however,
a characteristic feature of the synchronization by common noise. Indeed, even in the case of a vanishing coupling,
one observes phase locking by common noise, but the frequencies remain the natural ones (see in Fig.~\ref{fig4}
the curve corresponding to the case $\mu=0$). This is explained by the particular intermittent dynamics of the phase differences,
which has a form of long epochs of phase coincidence, interrupted with short phase slips, at which the phase difference
changes by $2\pi$, see Fig.~\ref{fig7}. In this picture, a finite difference of frequencies may coexist with an almost perfect
phase synchrony, provided the slips are very short.

%%%%%%%%%%%%%%%%%%%%%%%%%%%%%%%%%%%%%%%%%%%%%%%%%%%%%%%%%
%%%%%%%%%%%%%%%%%%%%%%%%%%%%%%%%%%%%%%%%%%%%%%%%%%%%%%%%%

The qualitative picture of the slip-mediated phase difference dynamics is valid also for non-zero values of
coupling, provided
that the synchronizing effect of noise is stronger than the repulsion due to the coupling. The only difference is that now
the frequency of slips is smaller or larger compared to the coupling-free case, for attractive or
repulsive coupling, respectively.
This is clearly seen in Fig.~\ref{fig7}, where the cases of repulsive, vanishing, and attractive couplings are depicted.
For a very strong value of repulsive coupling, the synchronous state is no more attractive, and one observes the slip-free dynamics
of the phase difference. Noticeably, slips are observed (though rarely) for strong attractive coupling as well. Thus, common noise prevents
complete frequency entrainment for non-identical rotators.

%%%%%%%%%%%%%%%%%%%%%%%%%%%%%%%%%%%%%%%%%%%%%%%%%%%%%%%%%
\begin{figure}[!t]
\centering
\includegraphics[width=0.95\columnwidth]%
 {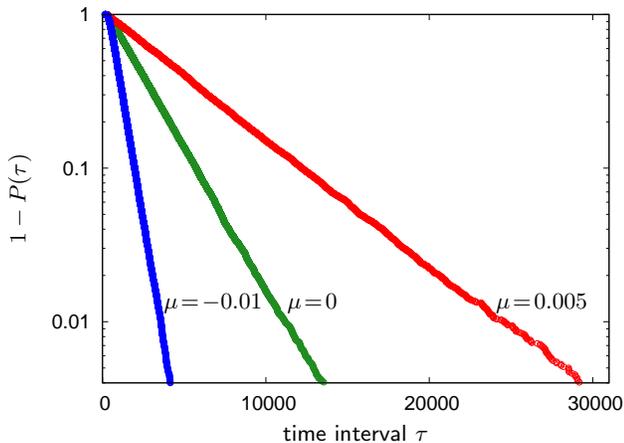}
\caption{(Color online)
Statistics of the time intervals between the slips, for different values of coupling strength: $\mu=0.005$,
$\mu=0$, and $\mu=-0.01$. Parameters of the population of rotators are the same as in Fig.~\ref{fig3}.
}
  \label{fig8}
\end{figure}
%%%%%%%%%%%%%%%%%%%%%%%%%%%%%%%%%%%%%%%%%%%%%%%%%%%%%%%%%

In Fig.~\ref{fig8} we present the distribution functions of the time intervals between the slips,
for the cases shown in Fig.~\ref{fig7}. The distribution function $P(\tau)$ is defined as the probability
that the time interval between the slips is less than $\tau$. One can see that with a good accuracy,
this distribution is exponential: $P(\tau)\approx\exp[-\tau/\langle\tau\rangle]$, i.e.\ the statistics of the
slips is a Poissonian one.

% \clearpage
\section{Conclusion}
In conclusion, we developed in this paper a theory of synchronization of coupled active rotators
by common noise. Contrary to independent noisy forces acting on
different rotators that desynchronize them,
common white noise always facilitates synchrony and even can
overcome repulsive coupling. We studied two situations,
one of the identical rotators, and a system of rotators with a disorder in the
torque. For the identical rotators, a fully synchronous state
appears when the coupling strength exceeds the threshold value~\eqref{eq:thr}.
Below this threshold, partial synchrony with a nonvanishing
fluctuating order parameter is observed. For the nonidentical rotators,
the effect of common noise is twofold. For an attractive coupling,
this noise although enhances synchrony, makes it less perfect: there is no exact
frequency locking of rotators, rather the frequency differences become small but remain finite
(cf.\ Fig.~\ref{fig3}). For a repulsive coupling, an interference of two opposite
actions of noise and of  coupling leads to a
juxtaposition of phase locking with frequency anti-entrainment:
while the phases of the rotators most of the time nearly coincide, their frequencies are pushed
aside and their difference is larger than that of the natural ones. We explain this effect by an
intermittent nature
of the phase dynamics: the phase differences are most of the time small (modulo $2\pi$), but this
locking is interrupted by the slips, which for repulsive coupling are
more frequent than in the uncoupled case.

Comparison of the results for the active rotators model with those for the Kuramoto--Sakaguchi
system of coupled oscillators~\cite{Pimenova_etal-16,Goldobin_etal-17} shows that the basic
effects are similar in these setups. However, technically the analysis of the active rotators
is more involved. One needs to introduce proper transformed variables to obtain the main effects
analytically, in the leading order of large frequency of rotations.

The analysis in this paper has been performed for the overdamped case, where the rotators are described by a
one-dimensional model~\eqref{eq:ar}. This ensures that the noise of any intensity
synchronizes rotators, because the corresponding Lyapunov exponent can be only negative.
For rotators with inertia, this geometrical restriction does not hold,
and strong noise may result in a positive largest Lyapunov exponent, thus
desynchronizing the rotators~\cite{Pikovsky-84,*Pikovsky-84a,Goldobin-Pikovsky-05b,*Goldobin-Pikovsky-06b}. This setup is of a potential relevance for power grid networks
with slightly imbalanced generators. A common external noise here could be due, e.g.,
to large-scale wind intensity fluctuations in a farm of wind turbines.

\appendix
\section{Averaging over fast oscillations}
\label{sec:apa}

The Fokker--Planck equation, following from the stochastic differential
equations~\eqref{eq:I}--\eqref{eq:th}, for the probability distribution
function $W(I,\Psi,\theta,t)$, reads
\begin{align}
&\frac{\partial W}{\partial t} +
\frac{\partial}{\partial I}
\Bigg\{\Big[
(\mu-2\gamma)I - 2\gamma(1-a\cos\Psi)I^2\Big]W\Bigg\}
\nonumber\\%[5pt]
& +\frac{\partial}{\partial \Psi}\Big\{\nu_0 W\Big\}
%\nonumber\\[8pt]
%&
 +\frac{\partial}{\partial \theta}
\Bigg\{\Bigg[\nu\big(1-a\cos(\Psi+\theta)\big)
\nonumber\\%[5pt]
&\quad  +\frac{\mu\Big(a\big(\sin(\Psi+\theta)-\sin\Psi\big)-\sin\theta\Big)}{1-a\cos\Psi}
\Bigg]W\Bigg\}
\nonumber\\%[8pt]
&\quad-\sigma^2\hat{Q}^2W=0\,,
\label{eq1-18}
\end{align}
where operator $\hat{Q}(\cdot)$ is defined as
\begin{equation}
\begin{gathered}
\hat{Q}(\cdot)%\textstyle
\equiv \frac{\partial}{\partial I}\left(-\frac{a I\sin\Psi}{\sqrt{1-a^2}}(\cdot)\right)
 +\frac{\partial}{\partial\Psi}\left(\frac{1-a\cos\Psi}{\sqrt{1-a^2}}(\cdot)\right)
\\%[5pt]
%\textstyle
 +\frac{\partial}{\partial\theta}\left(\frac{a}{\sqrt{1-a^2}}\big(\cos\Psi-\cos(\Psi+\theta)\big)(\cdot)\right).
\end{gathered}
 \label{eq2-01}
\end{equation}

On the basis of this Fokker--Planck equation, one can perform a rigorous procedure of averaging
over the fast rotation of the phase $\Psi$ for the case of high natural frequencies. We employ the
condition that the basic frequency of oscillations $\nu_0$ is large compared to parameters $\mu$,
$\gamma$, and $\sigma^2$ (which all have the dimension of inverse time). Parameter $B$ is not
assumed to be small, so that the parameter $a$ is finite.
For vanishing $\mu/\nu_0$, $\gamma/\nu_0$, and $\sigma^2/\nu_0$, the probability density
distribution $W(I,\Psi,\theta,t)=(2\pi)^{-1}w(I,\theta,t)$ is uniform in $\Psi$.
The probability density $w(I,\theta,t)$ is governed by the Fokker--Planck equation~(\ref{eq2-01}) averaged over $\Psi$. There are two equivalent methods for averaging over ``fast'' variables, the Krylov--Bogoliubov method~\cite{Stratonovich} and the method of multiple scales~\cite{Bensoussan}. Applying the latter one to Eq.~(\ref{eq2-01}) (for rigorous explanations on the procedure see Refs.~\cite{Pimenova_etal-16,Goldobin_etal-17}), one obtains:
\begin{equation}
\begin{aligned}
&\frac{\partial w(I,\theta, t)}{\partial t}\\
 &+\frac{\partial}{\partial I}\Bigg\{\Bigg[(\mu - 2\gamma)I -2\gamma I^2
 +\frac{a^2\sigma^2}{2(1-a^2)}I\Bigg]
w(I,\theta,t)\Bigg\}\\
&
+\frac{\partial}{\partial\theta}\Bigg\{\Bigg[\nu
+\frac{a^2\sigma^2}{2(1-a^2)}\bigg)\sin\theta\Bigg]
w(I,\theta,t)\Bigg\}
\\%[5pt]
&\qquad
-\sigma^2\hat{Q}_1^2w-\sigma^2\hat{Q}_2^2w=0\,,
\end{aligned}
\label{eq2-02}
\end{equation}
where the averaging
\[
\frac{1}{2\pi}\int\limits_0^{2\pi}\mathrm{d}\Psi\,\hat{Q}^2w(I,\theta,t)
=\hat{Q}_1^2w(I,\theta,t)+\hat{Q}_2^2w(I,\theta,t)
\]
yields two operators
\begin{equation}
\hat{Q_1}(\cdot) \equiv \frac{a}{\sqrt{2(1-a^2)}}\left[\frac{\partial}{\partial I}\Big(-I(\cdot)\Big)+\frac{\partial}{\partial \theta}\Big(\sin\theta(\cdot)\Big)\right]
\label{eq2-03}
\end{equation}
and
\begin{equation}
\hat{Q_2}(\cdot) \equiv \frac{a}{\sqrt{2(1-a^2)}}\frac{\partial}{\partial \theta}\Big((1-\cos\theta)(\cdot)\Big)\,.
\label{eq2-04}
\end{equation}

%\section{Two noise signals}
Eq.~(\ref{eq2-02}) can be treated as the Fokker--Planck equation for the stochastic
system \eqref{eq:Iav}--\eqref{eq:thav} with two independent Gaussian white normalized
noise signals $\zeta_1(t)$ and $\zeta_2(t)$.

\section{Asymptotic behavior of the frequency difference for small mismatches}
\label{sec:apb}

For $\nu/\widetilde{\sigma}^2\to0$, one can simplify Eq.~(\ref{eq4-03}). Indeed, the function in
the argument of the exponential is multiplied by $\nu/\widetilde{\sigma}^2$ and can be neglected
in domains where this function is finite; where the function tends to $\pm\infty$, it is non-negligible, but one
can use an approximate expression for it, $\cot(\theta/2)\approx2/\theta+2/(\theta-2\pi)$. Hence,
\begin{align}
\langle\dot{\theta}\rangle&\approx 4\pi\widetilde{\sigma}^2\Bigg\{
\int\limits_0^{2\pi}\mathrm{d}\theta
\int\limits_\theta^{2\pi}\mathrm{d}\psi
 \frac{(1-\cos\psi)^m}
  {(1-\cos\theta)^{1+m}}
\nonumber\\%[5pt]
&\times\exp\left[
-\frac{\nu}{\widetilde{\sigma}^2}\left(\frac{1}{\theta}+\frac{1}{\theta-2\pi}
-\frac{1}{\psi}-\frac{1}{\psi-2\pi}\right)\right]\Bigg\}^{-1},
\label{eq4-04}
\end{align}
where $m\equiv\mu/(2\widetilde{\sigma}^2)$. Let us consider separately two cases: $m<-1/2$
and $m>-1/2$, for which the integral either diverges or converges near the zero-value points of
cosine function, if one drops the cutting exponential factor. In domains where the integral converges without the
exponential cutting factor, this factor can be neglected; in domains where the integral diverges without this
factor, the principal contribution to the integral is made by a small vicinity of the divergence point
and one can employ this fact to simplify calculations.

For $m<-1/2$ (where $1+m<1/2$ as well), one can make the
substitution $(1-\cos\psi)^m\to2^{-m}/\psi^{-2m}+2^{-m}/(\psi-2\pi)^{-2m}$ and calculate
\begin{align}
&\int\limits_0^{2\pi}\mathrm{d}\theta
\int\limits_\theta^{2\pi}\mathrm{d}\psi
 \frac{(1-\cos\psi)^m}
  {(1-\cos\theta)^{1+m}}
\nonumber\\%[5pt]
&\times\exp\left[
-\frac{\nu}{\widetilde{\sigma}^2}\left(\frac{1}{\theta}+\frac{1}{\theta-2\pi}
-\frac{1}{\psi}-\frac{1}{\psi-2\pi}\right)\right]
\nonumber\\%[5pt]
&\approx\int\limits_0^{2\pi}\frac{\mathrm{d}\theta}{(1-\cos\theta)^{m+1}}
\int\limits_0^{2\pi}\mathrm{d}\psi
\frac{(\psi-2\pi)^{2m}}{2^m}\mathrm{e}^{\frac{\nu}{\widetilde{\sigma}^2(\psi-2\pi)}}
\nonumber\\%[5pt]
&\approx\frac{2^{-m}\sqrt{\pi}\,\Gamma(-m-\frac12)}{\Gamma(-m)}
\left(\frac{\nu}{\widetilde{\sigma}^2}\right)^{2m+1}\frac{\Gamma(-2m-1)}{2^m}
\nonumber\\%[5pt]
&=\frac{2\sqrt{\pi}\,\Gamma(-m-\frac12)\,\Gamma(-2m-1)}{\Gamma(-m)}
 \left(\frac{\nu}{2\widetilde{\sigma}^2}\right)^{2m+1}\,,
\label{eq4-05}
\end{align}
where we used that
\[
\int_0^{2\pi}(1-\cos\theta)^n\mathrm{d}\theta=\frac{2^{n+1}\sqrt{\pi}\,\Gamma(n+1/2)}{\Gamma(n+1)}\,.
\]
For $m>-1/2$ (where $1+m>1/2$ as well), one can make substitution $1/(1-\cos\theta)^{m+1}\to2^{m+1}/\theta^{2(m+1)}$ and calculate
\begin{align}
&\int\limits_0^{2\pi}\mathrm{d}\theta
\int\limits_\theta^{2\pi}\mathrm{d}\psi
 \frac{(1-\cos\psi)^m}
  {(1-\cos\theta)^{1+m}}
\nonumber\\%[5pt]
&\times\exp\left[
-\frac{\nu}{\widetilde{\sigma}^2}\left(\frac{1}{\theta}+\frac{1}{\theta-2\pi}
-\frac{1}{\psi}-\frac{1}{\psi-2\pi}\right)\right]
\nonumber\\%[5pt]
&\approx\int\limits_0^{2\pi}\frac{2^{m+1}\mathrm{d}\theta}{\theta^{2(m+1)}}
\mathrm{e}^{-\frac{\nu}{\widetilde{\sigma}^2\theta}}
\int\limits_0^{2\pi}\mathrm{d}\psi\,(1-\cos\psi)^m
\nonumber\\%[5pt]
&\approx\frac{2^{m+1}\sqrt{\pi}\,\Gamma(m+\frac12)}{\Gamma(m+1)}
\left(\frac{\nu}{\widetilde{\sigma}^2}\right)^{-2m-1}2^{m+1}\Gamma(2m+1)
\nonumber\\%[5pt]
&=\frac{2\sqrt{\pi}\,\Gamma(m+\frac12)\,\Gamma(2m+1)}{\Gamma(m+1)}
 \left(\frac{\nu}{2\widetilde{\sigma}^2}\right)^{-2m-1}\,.
\label{eq4-06}
\end{align}
Combining Eqs.~(\ref{eq4-05})--(\ref{eq4-06}), one can find from Eq.~(\ref{eq4-04}):
\begin{align}
\langle\dot{\theta}\rangle&=
\frac{2\sqrt{\pi}\,\Gamma\big(|m+\frac12|+\frac12\big)\,\widetilde{\sigma}^2}
 {\Gamma\big(|m+\frac12|\big)\,\Gamma\big(|2m+1|\big)}
 \left(\frac{\nu}{2\widetilde{\sigma}^2}\right)^{|2m+1|}\,.
\label{eq4-07}
\end{align}

\section{Dynamics of system~\eqref{eq1-04}--\eqref{eq1-05} for large $\Omega_0$ and
strong noise ($\sigma^2\sim\Omega_0$)}
\label{sec:apc}
Let us consider the stochastic system~\eqref{eq1-04}--\eqref{eq1-05} for the case
of $\Omega_0\sim\sigma^2\gg\mu\sim\gamma\sim B$. In this case, the order
parameter $J$ is nearly constant on the characteristic time scale of variation
of $\Phi$ governed by Eq.~\eqref{eq1-05}. Hence, one can solve Eq.~\eqref{eq1-05}
for $\Phi$ assuming $J$ to be frozen. Eq.~\eqref{eq1-05} yields the Fokker--Planck
equation for $W(\Phi,t)$\,:
\[
\frac{\partial W}{\partial t} +\frac{\partial}{\partial\Phi}\Big[\Big(\Omega_0-B\frac{J+1/2}{\sqrt{J(1+J)}}\sin\Phi -\sigma^2\frac{\partial}{\partial\Phi}\Big)W\Big]=0\,.
\]
For a steady state solution, this equation can be integrated once with respect to $\Phi$\,;
\[
\bigg(\Omega_0-B\frac{J+1/2}{\sqrt{J(1+J)}}\sin\Phi -\sigma^2\frac{\partial}{\partial\Phi}\bigg)W=q\,,
\]
where $q$ is a constant. In the leading order in $B$, one
finds $W^{(0)}(\Phi)=(2\pi)^{-1}$ and $q^{(0)}=\Omega_0/(2\pi)$.
For the first-order in $B$ correction $W^{(1)}(\Phi)$ we obtain:
\[
\bigg(\Omega_0-\sigma^2\frac{\partial}{\partial\Phi}\bigg)W^{(1)}
 =q^{(1)}+B\frac{J+1/2}{\sqrt{J(1+J)}}\sin\Phi\,W^{(0)}.
\]
Thus, $q^{(1)}=0$ and $W^{(1)}(\Phi)=A_{11}\sin\Phi+A_{12}\cos\Phi$, $A_{11}=\Omega_0\sigma^{-2}A_{12}$,
\[
A_{12}=\frac{B}{2\pi}\frac{\sigma^2}{\Omega_0^2+\sigma^4}\frac{J+1/2}{\sqrt{J(1+J)}}\,.
\]
Now, one can calculate $\langle\cos\Phi\rangle$ according to the distribution $W(\Phi)$,
and average Eq.~\eqref{eq1-04}, rewritten as
\[
\frac{\mathrm{d}}{\mathrm{d}t}\ln{J}=\mu-2\gamma(1+J)+B\sqrt{\frac{1+J}{J}}\cos\Phi\,,
\]
over fast rotations of $\Phi$, and over time:
\[
0=\mu-2\gamma(1+\langle{J}\rangle) +\frac{B^2\sigma^2}{2(\Omega_0^2+\sigma^4)}\left(1+\frac12\langle{J^{-1}}\rangle\right).
\]
According to~\eqref{eq:ct}, to the leading order, $\langle{J}\rangle=\langle{I}\rangle$ and we obtain Eq.~\eqref{eq:lnoise}:
\[
\langle{I}\rangle\approx\frac{1}{2\gamma}\left(\mu
 +\frac{B^2\sigma^2}{2(\Omega_0^2+\sigma^4)}\Big(1+\frac12\langle{I^{-1}}\rangle\Big)\right)-1\,.
\]

\begin{acknowledgments}
Studies presented in Secs.~\ref{sec:be}, \ref{sec:op} and Appendices~\ref{sec:apa} and \ref{sec:apc} of this work,
conducted by A.V.D.\ and D.S.G.,  were supported by the Russian Science Foundation (grant No.\ 14-12-00090).
Studies presented in Sec.~\ref{sec:fr} and Appendix~\ref{sec:apb}, conducted by A.P.,
were supported by the Russian Science Foundation (grant No.\ 17-12-01534).
The paper was finalized during the visit supported by G-RISC (grant No.\ M-2017b-5).
\end{acknowledgments}
%
% \bibliography{nld-old,nld-current,%
% pap-ab,pap-ce,pap-fg,pap-hj,pap-kl,%
% pap-mn,pap-oq,pap-rs,pap-tz,%
% pik,books,n-stand,from_all_databases}
\def\cprime{$'$}

\end{document}